\documentclass{WileyMSP-template}

\usepackage{amsmath}
\usepackage[table]{xcolor}
\usepackage{cite}
\usepackage{makecell}

\begin{document}

	\pagestyle{fancy}
	\rhead{\includegraphics[width=2.5cm]{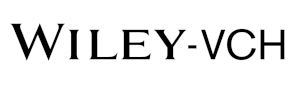}}

	\title{Molecular-optomechanical phonon laser}
	
	\maketitle

	
	\author{Bin Yin$^1$}
	\author{Jie Wang$^{2}$}
	\author{Qian Zhang$^1$}
	\author{Deng Wang$^1$}
	\author{Tian-Xiang Lu$^{3,*}$}
	\author{Hui Jing$^{1,2,\dagger}$}
	
	\begin{affiliations}
		$^1$Key Laboratory of Low-Dimensional Quantum Structures and Quantum Control of Ministry of Education, Department of Physics and Synergetic Innovation Center for Quantum Effects and Applications, Hunan Normal University, Changsha 410081, China\\
		$^2$Institute for Quantum Science and Technology, College of Science, NUDT, Changsha 410073, China\\		
		$^3$College of Physics and Electronic Information, Gannan Normal University, Ganzhou 341000, Jiangxi, China\\
		Email Address:
		*lu.tianxiang@foxmail.com, $^\dagger$jinghui73@foxmail.com

	\end{affiliations}
	
	
	\keywords{Phonon laser, Molecular optomechanics, Ultra-low threshold}
	
	\begin{abstract}
		
		Molecular cavity optomechanics (COM) leverages ultrastrong interactions between confined optical fields and high-frequency molecular vibration, providing a unique platform for exploring high-frequency phonon dynamics. In this work, we theoretically propose the use of a hybrid molecular COM system for realizing an ultra-low-threshold mid-infrared (MIR) phonon laser. Despite an optical quality factor of only $Q_a=100$, an ultra-low threshold power of $\mathrm{P}_{\mathrm{th}} = 17.5~\mathrm{nW}$ is achieved, enabled by giant single-photon optomechanical coupling and molecular collective effect. Moreover, the mechanical gain and threshold power can be further tuned by adjusting the distance between mirrors of the Fabry-P\'{e}rot cavity. Our findings establish the first direct connection between molecular COM and MIR phonon lasers, with potential applications in MIR acoustics and biomedical imaging.
		
	\end{abstract}
	

	\section{Introduction}
	Phonon lasers, the acoustic counterparts of optical lasers~\cite{Vahala2009,Pettit2019}, exhibit many similar characteristics~\cite{PhysRevA.90.053814} and play a crucial role in biomedical imaging~\cite{doi:10.1073/pnas.2413938121,10.1063/1.3459142}, quantum sensing~\cite{PhysRevApplied.16.044007,Pan.202400593,He:23}, and atmospheric monitoring~\cite{Pan.202400593,Xiao:23}. Over the past decade, phonon lasers have been experimentally realized on various platforms, such as trapped ions~\cite{Vahala2009}, levitated particles~\cite{kuang2023nonlinear,Xiao:23,He:23}, optical tweezers~\cite{xiao2024giant,Zhang:24}, and cavity optomechanical (COM) systems~\cite{Chafatinos2020,PhysRevLett.104.083901,zhang2018phonon,Wang:17}. These systems have demonstrated phonon laser with threshold powers at the microwatt level across frequency ranges from megahertz to gigahertz, offering remarkable performance characteristics, such as integrability~\cite{PhysRevLett.104.083901,zhang2018phonon}, frequency stability~\cite{Pan.202400593}, and potential high-speed signal processing capability~\cite{PhysRevLett.104.083901,zhang2018phonon}. Alongside these experimental breakthroughs, theoretical efforts have proposed novel mechanisms to further enhance phonon laser performance~\cite{PhysRevApplied.8.044020,PhysRevLett.113.053604,Wang_2023,PhysRevA.103.053501,PhysRevApplied.10.064037,Zhou:24,Huang_2023,lu2024quantum,Zhang_2018,Zhang_2023}, including exceptional point enhanced~\cite{PhysRevLett.113.053604,PhysRevApplied.8.044020,Wang_2023} and nonreciprocal~\cite{Huang_2023,PhysRevA.103.053501,PhysRevApplied.10.064037,Zhou:24,lu2024quantum} phonon lasers. Despite these advances, THz phonon lasers are still scarce and have fundamental limitations: especially the phonon lasers reported in COM are mainly limited to low-frequency operation, mainly at MHz, and still require relatively high threshold power. This limitation poses a critical barrier to their practical deployment, particularly in applications requiring terahertz-frequency acoustic waves, such as high-bandwidth acoustic filters~\cite{PhysRevLett.43.2012}, ultra-fast acoustic modulators~\cite{kittlaus2021electrically}, and strong thermal stability quantum circuits~\cite{yoon2024terahertz,sciadv.ado6240}. While achieving ultra-high frequency phonon lasing in COM remains challenging, the pursuit of such devices is of great significance.
	
	In recent years, a novel optomechanical platform—molecular optomechanics~\cite{benz2016single,esteban2022molecular,roelli2016molecular,ashrafi2019optomechanical}—has emerged. Bridging the fields of materials science and COM~\cite{RevModPhys.86.1391,10.1063/1.4896029,li2021cavity,doi:10.1126/science.1156032,jing2017high,jing2015optomechanically}, this platform exhibits several unique advantages that are unattainable in other optomechanical systems: (i) Leveraging the near-field enhancement effect of metal nanoparticles, it achieves the highest single-photon optomechanical coupling strength reported in the field to date~\cite{benz2016single,esteban2022molecular,roelli2016molecular}, along with natural spatial resolution at the single-molecule scale. (ii) By harnessing high-frequency intrinsic molecular vibrations~\cite{benz2016single,esteban2022molecular,roelli2016molecular}, it enables the highest mechanical frequencies among optomechanical systems, with the notable advantage of facile operation at room temperature. (iii) As a typical solid-state nanodevice, it offers compatibility with array configurations and large-scale integration. Within this framework, extensive theoretical and experimental efforts have explored phenomena~\cite{zhu2024autonomous,liu2018room,liu2017coupled,M+2025+59+73,PhysRevB.110.184306,chen2021continuous,xomalis2021detecting,zou2024amplifying,jakob2023giant,boehmke2024uncovering}, such as quantum heat engine~\cite{zhu2024autonomous}, precision measurement~\cite{liu2018room,liu2017coupled}, spring effects~\cite{jakob2023giant}, and low-frequency molecular vibrations~\cite{boehmke2024uncovering}. Molecular COM has great application prospects in areas such as single-molecule detection, terahertz sound waves, and high-frequency quantum acoustics. Importantly, the ultrahigh-frequency molecular vibrations in these systems offer a compelling platform for exploring phonon dynamics in the mid-infrared (MIR) regime. However, to our knowledge, the phonon coherence behavior in molecular COM has not been well developed, and the characteristics of the phonon laser have not been explored. At the same time, it has not been answered whether the high optical and mechanical dissipations of molecular COM seriously affect their behavior.  
	
	\begin{figure*}[t]
		\centering
		\includegraphics[width=0.8\linewidth]{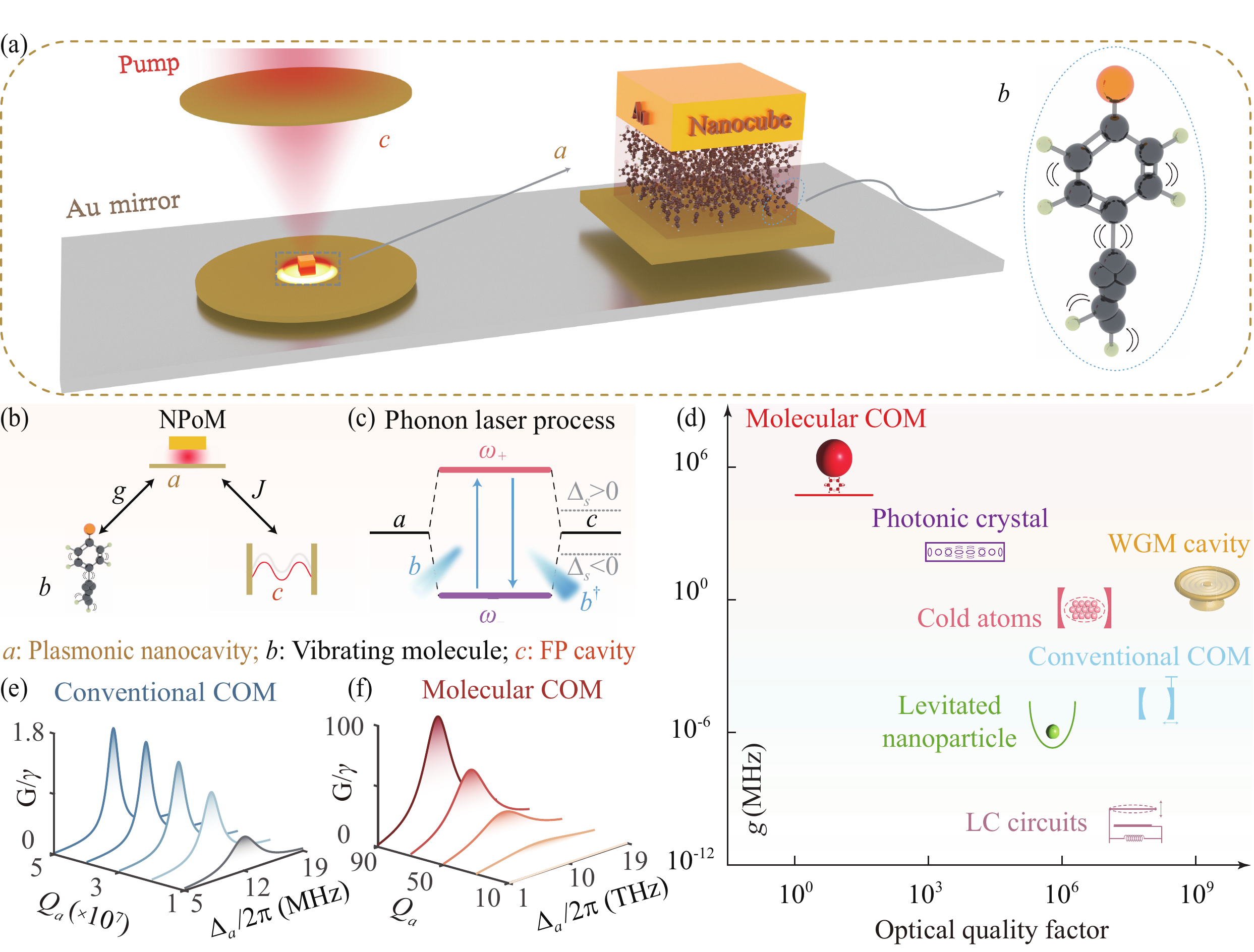}
		\caption{(a) Molecular COM system. The Au nanoblock is deposited on one mirror to form the nanoparticle-on-mirror configuration~\cite{HuXu+5163,li2024boosting,doi:10.1126/sciadv.adt9808,J-FLi,P.Alonso-González}, establishing a plasmonic nanocavity with resonance frequency $\omega_a$. A molecular ensemble with characteristic frequency $\omega_b$ is placed at the nanocavity's hotspot. These Au mirrors collectively constitute an FP cavity supporting an optical mode at frequency $\omega_c$. (b) Equivalent mode coupling. The plasmonic nanocavity interacts with molecules through optomechanical coupling $g$ and with the FP cavity via evanescent field $J$. (c) Phonon laser mechanism. (d) Comparison of optomechanical coupling strength $g$ and optical quality factor $Q$ characterizing different types of optomechanical systems~\cite{roelli2016molecular,verhagen2012quantum,murch2008observation,akahane2005fine,liu2011high,bagci2014optical,thompson2008strong,zhang2024miniature}. (e-f) The normalized mechanical gain $\mathrm{G}/\gamma$ as a function of the optical detuning $\Delta_a$ at the different quality factors $Q_a$. The parameters used here are $\mathrm{P}_\mathrm{l}=10\,\mathrm{mW}$ and $N=10^3$.} 
		\label{fig1}
	\end{figure*}
	
	Based on the significant challenges and unresolved issues mentioned above, we have developed a hybrid molecular COM system, utilizing the idea of coupling high-Q cavity with low-Q cavity~\cite{PhysRevLett.130.143601,shlesinger2023hybrid,zhang2024miniature,yinbin2025PRA}. We theoretically investigate phonon laser within a hybrid molecular COM system by integrating an Au nanoblock and a molecular ensemble into a Fabry-P\'erot (FP) cavity. We found that: (i) An exceptionally low threshold power of $\mathrm{P}_{\mathrm{th}} = 17.5\,\mathrm{nW}$ can be achieved even with a low optical quality factor of $Q_a = 100$. (ii)  The FP cavity enables precise control over phonon laser behavior by tuning the distance between mirrors. (iii) This system supports MIR phonon laser $\omega_b/2\pi=20\,\mathrm{THz}$, holding significant promise for applications such as laser-based sensing, biomedicine~\cite{doi:10.1073/pnas.2413938121,10.1063/1.3459142} and MIR acoustics. These results underscore the potential of hybrid molecular COM systems as versatile platforms for coherent phonon amplification, paving the way for novel advancements in the molecular COM systems and related acoustic technologies.
	
	\section{Theoretical model}\label{sec2}
	As illustrated in Figure~\ref{fig1}a, we theoretically consider a hybrid molecular COM system by integrating an Au nanoblock, a molecular ensemble, and two highly reflective Au mirrors. The Au nanoblock is deposited on one mirror to form the nanoparticle-on-mirror configuration~\cite{HuXu+5163,li2024boosting}, establishing a plasmonic nanocavity. This nanocavity supports an optical mode with the resonance frequency $\omega_a$ and quality factor $Q_a$ driven by a pump laser at frequency $\omega_l$. In parallel, another adjustable mirror is combined with the fixed mirror to form an FP cavity. This FP cavity also supports an optical mode with resonance frequency $\omega_c$ and quality factor $Q_c$, which couples to the plasmonic nanocavity via the evanescent field with a coupling strength $J$~\cite{shlesinger2023hybrid,shlesinger2021integrated}, as depicted in Figure~\ref{fig1}b. Moreover, the molecules are precisely localized at the hotspot of the plasmonic nanocavity, and due to the vibration of the molecules with a resonance frequency $\omega_b$, the resonance frequency of the cavity field changes, enabling optomechanical coupling $g=\omega_{a}R_{b}\sqrt{\hbar/2\omega_{b}}/\epsilon_{0}V_{m}$ with the cavity mode~\cite{benz2016single,esteban2022molecular,roelli2016molecular}, where $R_{b}$ is the Raman polarizability and $V_{m}$ is the effective mode volume of the cavity. NPoM structures are compatible with a variety of nanoparticles and, together with cavity gaps, modulate cavity resonance frequencies $\omega_a$ and effective mode volumes $V_{m}$. Moreover, molecular optomechanics achieves an "equivalent breakthrough" in the diffraction limit via molecular localization and near-field optics, where the effective coupling volume is determined by the molecular spatial volume~\cite{jakob2023giant,benz2016single,roelli2016molecular}. Integrating surface plasmons or nanophotonic structures localizes optical fields to the sub-diffraction nanoscale, reducing the mode volume to less than one thousandth of $V_{m}\sim\lambda^{3}$, and significantly enhancing photon-molecule coupling (more details see the Appendix.~\ref{A1}). Therefore, the Hamiltonian of the system can be expressed as:
	\begin{eqnarray}
		\begin{aligned}
			H &= \hbar\omega_a a^{\dagger}a+\hbar\omega_c c^{\dagger}c+\sum_{i}^{N}\hbar\omega_{b}b_{i}^{\dagger}b_{i}+\hbar J(a^{\dagger}c+ac^{\dagger}) \\ 
			&~~~-\sum_{i}^{N}\hbar ga^{\dagger}a(b_{i}^{\dagger}+b_{i})+i\hbar\epsilon_{l}(a^{\dagger}e^{-i\omega_l t}-ae^{i\omega_l t}),\end{aligned}
		\label{1}
	\end{eqnarray}
	where $a\,(a^{\dagger}$), $c\,(c^{\dagger}$), and $b_i\,(b^{\dagger}_i)$ are the annihilation (creation) operators for the plasmonic nanocavity optical mode, FP cavity optical mode, and $i$-th individual molecular vibrational mode, respectively. $\epsilon_{l}=\sqrt{{2\kappa_a\mathrm{P}_\mathrm{l}}/{\hbar\omega_l}}$ is the driving amplitude with input power $\mathrm{P}_\mathrm{l}$.
	Here, we discussed a large number of molecules, i.e., $N$$\,\gg\,$$1$, which is always feasible in experiments~\cite{boehmke2024uncovering,shlesinger2023hybrid}. Our motivation is to analyze the influence of molecular number $N$ on the performance of this phonon laser. To this end, one defines the collective effect operator of molecules $B=\sum_{i=1}^{N}b_{i}/\sqrt{N}$~\cite{zhang2020optomechanical,PhysRevB.110.184306,zou2024amplifying}, and in the rotating framework at frequency $\omega_l$, the Hamiltonian of the system can be simplified as:
	\begin{eqnarray}
		\begin{aligned}
			H =&-\hbar\Delta_a a^{\dagger}a-\hbar\Delta_c c^{\dagger}c+\hbar\omega_{b}B^{\dagger}B+\hbar J(a^{\dagger}c+ac^{\dagger}) \\ &-\hbar g_{N}a^{\dagger}a(B^{\dagger}+B)+i\hbar\epsilon_{l}(a^{\dagger}-a),\end{aligned}
		\label{2}
	\end{eqnarray}
	where $\Delta_{a,c}=\omega_{l}-\omega_{a,c}$. $g_N=g\sqrt{N}$ is optomechanical coupling caused by collective effect~\cite{PhysRevB.110.184306,zou2024amplifying}. As confirmed in the experiment~\cite{PhysRevLett.104.083901}, we focus on the average behavior and ignore the quantum noise term, thus obtaining the steady-state mean values:
	\begin{eqnarray}
		\begin{aligned}
			a_s= &\frac{-iJc_s+\epsilon_{l}}{-i\Delta_a+\kappa_{a}-ig_{N}(B_s^*+B_s)}, B_s=\frac{ig_{N}|a_s|^{2}}{i\omega_{b}+\gamma},c_s=\frac{-iJa_s}{-i\Delta_c+\kappa_{c}}.\end{aligned}
		\label{3}
	\end{eqnarray} 
	$\kappa_a$, $\kappa_c$, and $\gamma$ are the decay rates of the plasmonic nanocavity optical mode, FP cavity optical mode, and molecular collective mode, respectively. To analyze the phonon laser behavior in this hybrid molecular COM system, we derive a simplified Hamiltonian using supermode operators $a_{\pm}=(a\pm c)/\sqrt{2}$ and rotating-wave approximation (more details see the Appendix.~\ref{A2})~\cite{PhysRevLett.104.083901,PhysRevApplied.10.064037}. And we define the ladder operator $p=a_{-}^{\dagger}a_{+}$ and population inversion operator $\delta n=a_{+}^{\dagger}a_{+}-a_{-}^{\dagger}a_{-}$ for optical supermode~\cite{PhysRevLett.104.083901,Wang:17,zhang2018phonon}, thus further obtaining motion equations of the system,
	
	\begin{eqnarray}
		\begin{aligned}
			\dot{B}=&-(i\omega_{b}+\gamma)B+\frac{ig_{N}}{2}p, \\ \dot{p}=&-2(iJ+\kappa)p+\frac{i}{2}(\Delta_{s}-g_{N}B)\delta n+\frac{\epsilon_{l}}{\sqrt{2}}(a_{-}^{\dagger}+a_{+}),\end{aligned}
		\label{4}
	\end{eqnarray} 
	where $\kappa=(\kappa_a+\kappa_c)/2$. By solving Eq.~(\ref{4}), we can easily obtain the mechanical gain $\mathrm{G}$ (more details see the Appendix.~\ref{A3}),
	\begin{eqnarray}
		\begin{aligned}
			\mathrm{G}=&\frac{g_{N}^{2}\kappa\delta n}{2(2J-\omega_{m})^{2}+8\kappa^{2}}+\frac{\epsilon_{l}^{2}g_{N}^{2}(\omega_{m}-2J)(\Delta_{a}+\Delta_{c})\kappa}{[4(2J-\omega_{m})^{2}+16\kappa^{2}][D^{2}+(\Delta_{a}+\Delta_{c})^{2}\kappa^{2}]},\end{aligned}
		\label{5}
	\end{eqnarray}
	with
	\begin{eqnarray}
		\begin{aligned}
			D=&J^{2}+\kappa^{2}-\Delta_{a}\Delta_{c}+\frac{g_{N}^{2}B^{\dagger}B-2\Delta_{s}g_{N}Re(B)}{4}, \\ \delta n=&\frac{\epsilon_{l}^{2}[2J\Delta_{c}-Jg_{N}\mathrm{Re}(B)-\kappa g_{N}\mathrm{Im}(B)]}{D^{2}+\kappa^{2}(\Delta_{a}+\Delta_{c})^{2}}.\end{aligned}
		\label{6}
	\end{eqnarray}
	Next, we theoretically study the behavioral characteristics of phonon laser in this system.
	
	\section{Results and discussion}\label{sec3}
	To demonstrate the experimental feasibility and practical applicability of the proposed phonon laser, we selected feasible experimental parameters in the numerical simulation process: The conventional COM system~\cite{PhysRevLett.113.053604,peng2014parity}, i.e., $\omega_a=\omega_c=193\,\mathrm{THz}$, $\omega_b/2\pi=24\,\mathrm{MHz}$, $Q_a=Q_c=3$$\,\times\,$$10^7$, $\gamma=0.24\,\mathrm{MHz}$, $J=0.5\omega_b$, $g=465.3\,\mathrm{Hz}$. And the hybrid molecular COM system (more details see the Appendix.~\ref{A1})~\cite{shlesinger2023hybrid,pan2019elucidating,esteban2022molecular}, i.e., $\omega_a/2\pi=\omega_c/2\pi=460\,\mathrm{THz}$, $\omega_b/2\pi=20\,\mathrm{THz}$, $Q_a=100$, $Q_c=10^4$,  $\gamma/2\pi=0.01\,\mathrm{THz}$, $J=0.5\omega_b$, $g/2\pi=70\,\mathrm{GHz}$. 
	\begin{figure}[t]
		\centering
		\includegraphics[width=0.7\linewidth]{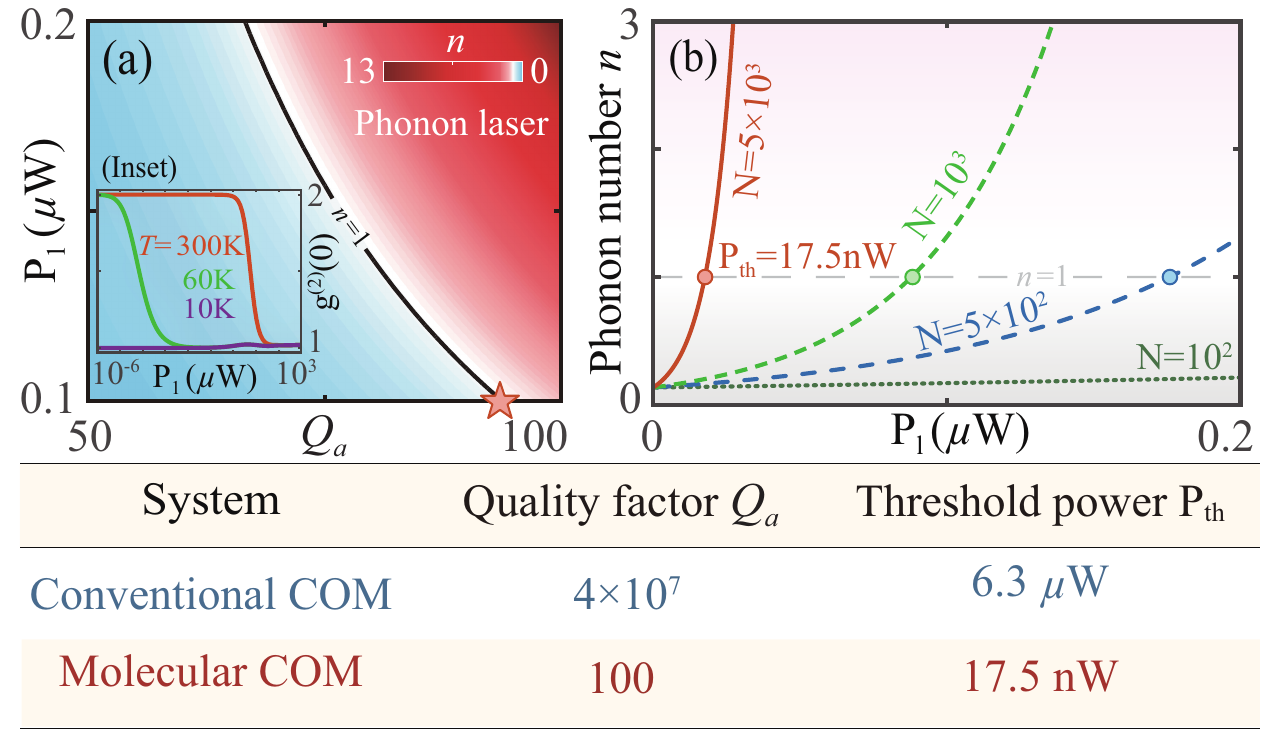}
		\caption{(a) The stimulated emitted phonon number $n$ versus the quality factor $Q_a$ and the driving power $\mathrm{P}_\mathrm{l}$. (Inset) The second-order correlation $g^{(2)}(0)$ versus $\mathrm{P}_\mathrm{l}$ at different temperatures. (b) $n$ versus $\mathrm{P}_\mathrm{l}$ at different numbers of molecules $N$. The parameters used here are $N=10^3$ (a) and $Q_a=12.5$ (Inset).} 
		\label{fig2}
	\end{figure}
	\subsection{Mechanical gain and phonon-lasing threshold}
	In this section, we theoretically analyze the mechanical gain in both conventional and hybrid molecular COM systems. As shown in Figure~\ref{fig1}e-f, the normalized mechanical gain $G/\gamma$ is plotted as a function of the quality factor $Q_a$ and the optical detuning $\Delta_a$, under a fixed driving power. For the conventional COM system [see Figure~\ref{fig1}e], the mechanical gain $G$ increases with the optical quality factor $Q_a$. Under suitable parameters, the system can enter the phonon lasing regime, characterized by $\mathrm{G}/\gamma>1$. At the optimal detuning $\Delta_a=0.5\omega_b$, 
	the mechanical gain reaches a peak value of $\mathrm{G}/\gamma=1.58$. Similarly, in the hybrid molecular COM system [see Figure~\ref{fig1}f], the system also accesses the phonon lasing regime, achieving a significantly higher mechanical gain. At the optimal detuning $\Delta_a=0.5\omega_b$, the gain can reach as high as $\mathrm{G}/\gamma=90$. The physical mechanism underlying the observed phonon lasing behavior can be understood from the energy-level diagram [see Figure~\ref{fig1}c]. When the frequency difference between these supermodes matches the mechanical mode frequency, i.e., $\omega_+-\omega_-=\omega_b$, and the driving field is resonant with the higher-energy supermode, population inversion is achieved. Specifically, the driving laser populates the higher-energy supermode with sufficient photons. This enables a stimulated transition to the lower-energy supermode, during which a phonon is emitted. As a result, higher-energy supermode photons are converted into lower-energy supermode photons accompanied by coherent phonon generation, thus amplifying the mechanical motion~\cite{PhysRevLett.104.083901,Wang:17,zhang2018phonon}.
	
	Once the mechanical gain $\mathrm{G}$ is obtained, the stimulated emitted phonon number $n$ can be calculated, $n=\mathrm{exp}[2(\mathrm{G}-\gamma)/\gamma]$,
	which characterizes the performance of the phonon laser~\cite{PhysRevLett.104.083901}. Subsequently, under the threshold condition of the phonon laser $n=1$ (i.e., $\mathrm{G}=\gamma$)~\cite{PhysRevLett.104.083901}, we derive the threshold power as follows:
	\begin{eqnarray}
		\begin{aligned}
			\mathrm{P}_{\mathrm{th}}\approx \frac{2\hbar\kappa\gamma\omega_{l}(J^{2}+\kappa^{2}-\Delta_{a}\Delta_{c})^{2}+2\hbar\kappa^{3}\gamma\omega_{l}(\Delta_{a}+\Delta_{c})^{2}}{\kappa_{a}Jg_{N}^{2}\Delta_{c}},\end{aligned}
		\label{7}
	\end{eqnarray}
	in which we have used $|B_{s}|^{2}\ll1$.
	
	As shown in Figure~\ref{fig2}, we theoretically analyze the stimulated phonon number $n$ and threshold power $\mathrm{P}_\mathrm{l}$ in the proposed hybrid molecular COM systems. For the hybrid molecular COM system, it is observed that phonon laser, i.e., $n>1$, can be achieved under appropriate parameter conditions in Figure~\ref{fig2}a. When $Q_a=100$, the threshold power can be reduced to $\mathrm{P}_{\mathrm{th}}=88\,\mathrm{n W}$ [see the pentagram in Figure~\ref{fig2}a], which is nearly two orders of magnitude lower than the conventional COM system. This advantage stems from its inherently strong single-photon optomechanical coupling and the molecular collective effect, which together support coherent phonon emission at ultra-low quality factors and substantially lower threshold power. This feature is particularly promising for practical applications where non-invasive phonon-based diagnostics or therapies are desired, such as in biomedical sensing and therapy~\cite{doi:10.1073/pnas.2413938121,10.1063/1.3459142}. 
	
	In addition, the influence of molecular collective effect is illustrated in Figure~\ref{fig2}b. The results indicate that an increase in the number of molecules enhances the effective optomechanical coupling, leading to a higher stimulated emitted phonon number $n$. Concurrently, the threshold power for the phonon laser decreases with a greater number of molecules. Notably, when the number of molecules reaches $N=5\times10^3$, the threshold power can be reduced to as low as $\mathrm{P}_{\mathrm{th}}=17.5\,\mathrm{nW}$. More noteworthy, due to the characteristics of molecules $\omega_b/2\pi=20\,\mathrm{THz}$~\cite{benz2016single,esteban2022molecular,roelli2016molecular}, this device can achieve the MIR-frequency phonon laser, which is significantly different from conventional COM system phonon laser, $\omega_b/2\pi=24\,\mathrm{MHz}$. High-frequency phonon lasers are distinguished by their exceptional resolution~\cite{RevModPhys.86.1391,10.1063/1.1499745}, robust penetration capabilities~\cite{https://doi.org/10.1002/lpor.201000011}, and low noise levels~\cite{teufel2011sideband,chan2011laser}.
	\subsection{Second-order quantum coherence of phonon lasing}
	
	\begin{figure*}[t]
		\centering
		\includegraphics[width=0.8\linewidth]{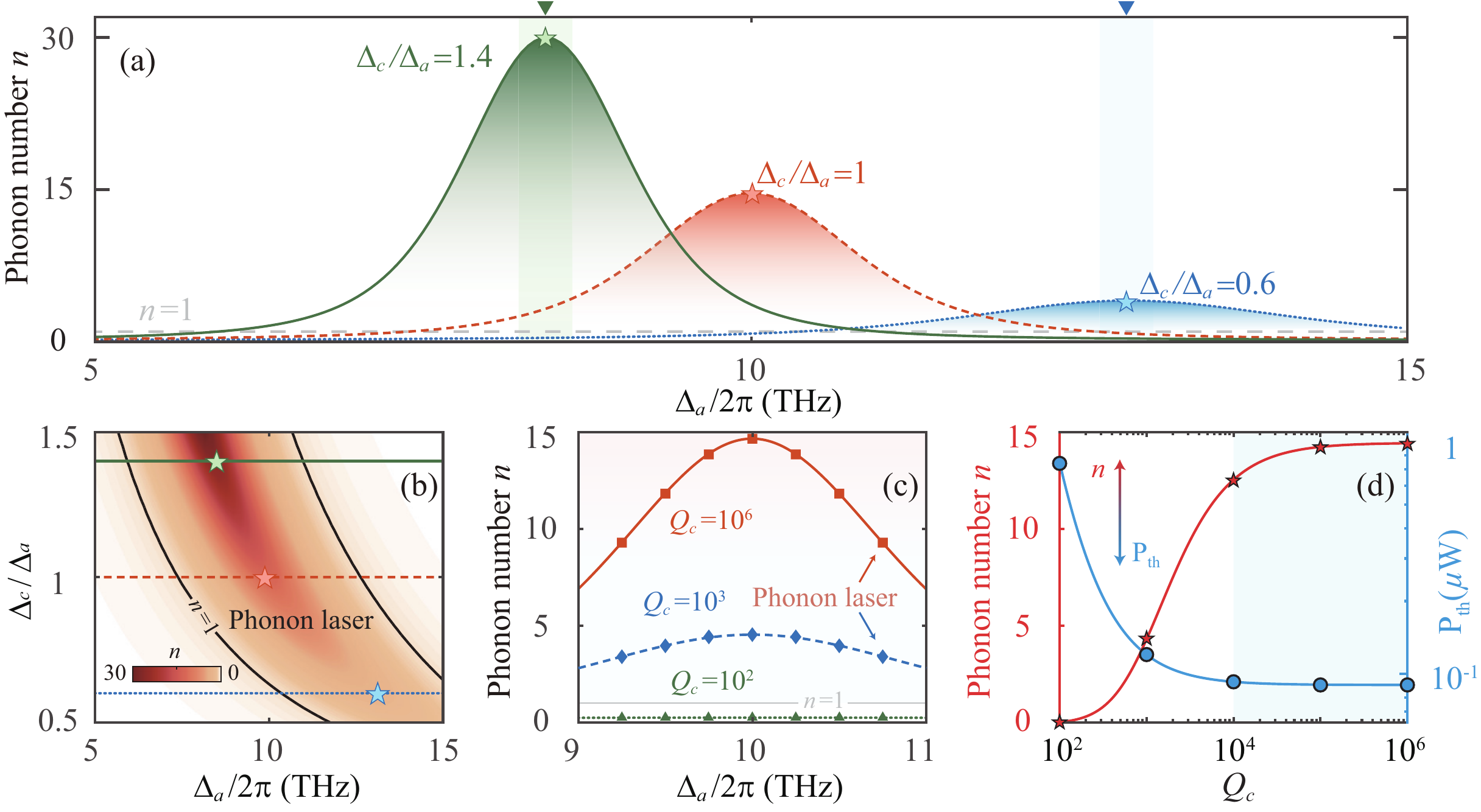}
		\caption{(a-b) The stimulated emitted phonon number $n$ as a function of optical detuning $\Delta_a$ and $\Delta_c$. (c) $n$ as a function of optical detuning $\Delta_a$ under different quality factors $Q_c$. (d) $n$ and threshold power $\mathrm{P}_{\mathrm{th}}$ as a function of quality factor $Q_c$. We choose $\mathrm{P}_\mathrm{l}=0.2\,\mathrm{\mu W}$ and $N=10^3$ in all these figures.} 
		\label{fig3}
	\end{figure*}
	Meanwhile, we also theoretically explored the statistical characteristics of phonons. By adopting the same approach used for photons, we calculated the normalized equal-time second-order phonon autocorrelation functions~\cite{PhysRevA.90.053814,doi:10.1126/sciadv.adv9984}, defined as $g^{(2)}(0)=\langle B^{\dagger}(t)B^{\dagger}(t)B(t)B(t)\rangle/\langle B^{\dagger}(t)B(t)\rangle^{2}$. By leveraging the small-fluctuation approximation (more details see the Appendix.~\ref{A4}), we can express the second-order coherence degree as follows
	\begin{eqnarray} 
		\begin{aligned}
			g^{(2)}(0)=&\frac{|B_{s}|^{4}+2Re[B_{s}^{*2}\langle\beta^{\dagger}(t) \beta(t)\rangle]}{[|B_{s}|^{2}+\langle\beta(t)\beta(t)\rangle]^{2}}+\frac{4|B_{s}|^{2}\langle\beta^{\dagger}(t)\beta(t)\rangle+\langle\beta^{\dagger}(t)\beta^{\dagger}(t)\beta(t)\beta(t)\rangle}{[|B_{s}|^{2}+\langle\beta(t)\beta(t)\rangle]^{2}}.
		\end{aligned}
		\label{8}
	\end{eqnarray}
	Here, we study the zero-time-delay second-order phonon autocorrelation function $g^{(2)}(0)$ at different temperatures, as shown in the inset of Figure~\ref{fig2}a. At room temperature $\,\mathrm{T}=300\,\mathrm{K}$, $g^{(2)}(0)=2$ under low driving power $\mathrm{P}_\mathrm{l}$, manifesting the characteristic thermal statistical properties. As $\mathrm{P}_\mathrm{l}$ increases, $g^{(2)}(0)$ approaches 1, indicating that the phonon dynamics transitions from the thermal state to the coherent state. This behavior originates from the extremely high molecular vibrational frequency $\omega_b/2\pi=20\,\mathrm{THz}$, which results in a low average thermal phonon occupation number in the system itself at room temperature.
	\\When the temperature is decreased to $\,\mathrm{T}=60\,\mathrm{K}$, $g^{(2)}(0)=1$ can be more easily achieved, as thermal noise will be significantly further suppressed at low temperatures. Upon further cooling to $\,\mathrm{T}=10\,\mathrm{K}$, thermal noise becomes nearly negligible, and the phonons approach an ideal coherent state. This is consistent with the mechanism of noise suppression mentioned in reference~\cite{benz2016single}, where cooling to $\,\mathrm{T}=10\,\mathrm{K}$ effectively suppresses thermal noise and allows an accurate measurement of the non-thermal vibrational population caused by optomechanical coupling.

	\subsection{Piezoelectric-controlled phonon laser}
	Building upon our previous analysis of the plasmonic nanocavity's parametric influence on the phonon laser, we now extend our investigation to theoretically explore the effects of the FP cavity. Firstly, the characteristics of this phonon laser can also be modulated by varying the distance between mirrors. In the experiment~\cite{shlesinger2023hybrid}, the distance between two mirrors can be precisely controlled by a piezoelectric translation stage, supporting nanometer-level precision scanning and continuously adjusting the cavity length, thereby adjusting the optical detuning, i.e., modifying the detuning parameter $\Delta_c$. As shown in  Figure~\ref{fig3}a, we further examine the variation of $n$ for three specific detuning values $\Delta_c$. The results reveal that $n$ increases with a redshift when $\Delta_c/\Delta_a>1$, whereas it decreases with a blueshift when $\Delta_c/\Delta_a<1$~\cite{PhysRevApplied.10.064037,lu2024quantum}. By selecting an appropriate detuning $\Delta_c$, the stimulated emitted phonon number can be changed from $n<1$ to $n>1$. In Figure~\ref{fig3}b, the stimulated emitted phonon number $n$ is plotted as a function of the $\Delta_a$ and $\Delta_c$. This reveals that tuning the cavity length can either enhance or suppress phonon emission and induce a frequency shift. These findings demonstrate that the phonon laser can be effectively switched. The physical mechanism can be explained as follows. From the energy levels in Figure~\ref{fig1}c, the variation in the resonant frequency $\omega_c$ induces corresponding shifts in the frequencies of both supermodes. To maintain the optimal pumping condition where $\omega_{l}=\omega_{+}$, the pump frequency must undergo either a red or blue shift. Consequently, the detuning difference $\Delta_c$ causes frequency shift and influences the steady-state photon number $|a_s|^2$, subsequently modifying the effective optomechanical coupling strength, thereby achieving modulation of phonon laser characteristics~\cite{PhysRevA.103.053501,PhysRevApplied.10.064037,lu2024quantum,Zhou:24}. 
	
	Moreover, we can improve the performance of phonon lasers by optimizing the quality factor of the FP cavity. Figure~\ref{fig3}c shows the stimulated phonon number $n$ as a function of optical detuning $\Delta_{a}$ under different quality factors $Q_c$. A clear trend emerges: higher $Q_c$ leads to larger $n$. Figure~\ref{fig3}d further shows the dependence of $n$ and the threshold power $\mathrm{P}_{\mathrm{th}}$ on $Q_c$ at a fixed detuning $\Delta_a = 0.5\omega_b$. As $Q_c$ increases, phonon emission is enhanced and $\mathrm{P}_{\mathrm{th}}$ decreases accordingly. However, beyond a critical value (e.g., $Q_c=10^4$), this trend saturates [see Figure~\ref{fig3}d], as further increases in $Q_c$ do not significantly enhance the effective optomechanical interaction. Notably, an efficient phonon laser can already be achieved without requiring an excessively high cavity quality, i.e., $Q_c=10^4$, significantly easing experimental demands and enhancing practical viability. Based on these, our device provides a straightforward and practical approach. Through systematic adjustment of the quality factor $Q_c$ and the distance between mirrors, we can achieve comprehensive control over the characteristics of the phonon laser.
	
	\subsection{Advantage and experimental feasibility}
	
	\smallskip
	
	\begin{table}[htbp]
		\caption{The characteristics of phonon lasers in different experimental systems}
		\label{Table-1}
		\centering
		\setlength{\tabcolsep}{25pt}  
		\begin{tabular}{cccccc}
			\rowcolor{red!12}
			\hline
			System & $\omega_b/2\pi$ & $\mathrm{P}_{\mathrm{th}}$  & $Q_a$ & Gain \\
			\rowcolor{red!3}
			\hline
			WGM~\cite{PhysRevLett.104.083901} &  $23.4\times10^6$ & $7\,\mathrm{\mu W}$ & $4\times10^7$  &  \\\rowcolor{red!3}
			WGM~\cite{zhang2018phonon} & $17.4\times10^6$ & $0.2\,\mathrm{mW}$ & $6.33\times10^7$  &  \\\rowcolor{red!3}
			FP~\cite{PhysRevLett.124.053604} &  $1.2\times10^6$ & $3\,\mathrm{mW}$  & / &  \\\rowcolor{red!3}
			Photonic crystal~\cite{Cui:21} &  $5.91\times10^9$ & $31\,\mathrm{\mu W}$  & $10^4$  &  \\\rowcolor{red!3}
			Electromechanics~\cite{PhysRevLett.110.127202} & $2.5\times10^6$ & / & / &  \\\rowcolor{red!3}
			Nano sphere~\cite{kuang2023nonlinear} &  $14.4\times10^3$ & $0.39\,\mathrm{mW}$  & $10^6$ & Y  \\\rowcolor{red!3}
			Nanosphere~\cite{xiao2024giant} &  $10.1\times10^3$ & $30\,\mathrm{mW}$  & $10^6$ & Y  \\\rowcolor{red!3}
			Polariton~\cite{Chafatinos2020} & $20\times10^9$ & $19\,\mathrm{mW}$  & / &  \\
			\hline
			\rowcolor{red!8}
			This work & $20\times10^{12}$ & $17.5\,\mathrm{nW}$  &  &  \\
			\hline
		\end{tabular}
	\end{table}
	
	We highlight three noteworthy and novel results of our work. (i) To the best of our knowledge, the phonon laser properties in molecular optomechanical systems have not been previously explored. We systematically studied phonon lasers in molecular COM systems for the first time, and specifically investigated their characteristics and coherence behavior. (ii) In addition to theoretical discussions, this study aims to develop a novel phonon laser micro nano device and overcome a key technical bottleneck: the limited application of low-frequency phonon lasers in the high-frequency region. It is worth noting that our strategy successfully breaks through this limitation and demonstrates an ultra-high frequency MIR phonon laser. More broadly speaking, this work provides a new approach to address current limitations and opens up new directions for high-frequency quantum acoustics. (iii) In various application scenarios of phonon lasers, the threshold is the core determining factor. However, in response to the bottleneck of high threshold values in practical applications, this study successfully achieves the $\mathrm{P}_{\mathrm{th}}=17.5\,\mathrm{nW}$ level ultra-low threshold phonon laser. More interestingly, to our knowledge, our scheme can achieve the lowest threshold level and highest operating frequency among phonon laser devices known in the field of cavity optomechanics without any additional auxiliary means. In the long run, the scheme proposed in this study is undoubtedly a powerful platform for further breakthroughs in the performance of phonon lasers.
		\\
		In addition, we will highlight the originality and advantages of this work by comparing it in detail with previous work. Previous studies have explored phonon lasers in COM, utilizing techniques such as rotation, squeezing light, phase control, and gain media, as well as various methods in cavity magnetic systems [see Table~\ref{Table-1}]. However, our work has achieved an unprecedented level that none of the above can achieve using the simplest elements. We can achieve the ultra-low threshold MIR phonon laser, i.e., $\mathrm{P}_{\mathrm{th}}=17.5\,\mathrm{nW}$ and $\omega_b/2\pi=20\,\mathrm{THz}$, without any nonlinearity or gain. Compared with traditional optomechanical systems, i.e., $\mathrm{P}_{\mathrm{th}}=6.3\,\mathrm{\mu W}$ and $\omega_b$$\,\sim\,$$\,\mathrm{MHz}$, the threshold has decreased by 2-3 orders of magnitude, and the laser frequency has increased by 5-6 orders of magnitude. That is to say, our scheme does not require any additional auxiliary means to achieve the lowest threshold level and the highest frequency level of device operation, as far as we know in the field of cavity optomechanics. From a longer-term perspective, the scheme proposed in this study is definitely a groundbreaking universal path in the threshold and operating frequency of phonon lasers.
	
	Finally, we focus on the feasibility analysis of the experiment. The core of this physical model lies in the NPoM structure~\cite{HuXu+5163,li2024boosting,doi:10.1126/sciadv.adt9808}. The NPoM structure has universality. It can adapt to various shapes of nanoparticles, such as nanospheres~\cite{benz2016single,jakob2023giant,boehmke2024uncovering,zhang2024miniature}, nanoblocks~\cite{shlesinger2023hybrid}, and metal nanoantennas~\cite{Q.Ma,F.Borghese}. It can also adapt to various gap materials such as molecules~\cite{benz2016single,jakob2023giant,boehmke2024uncovering}, two-dimensional materials~\cite{T.G.Habteyes,J-FWang}, rare-earth ions~\cite{H-XXu}, quantum dots~\cite{Q.Ma,F.Borghese}, and so on. In typical molecular COM systems, two key elements that need to be emphasized are ensuring atomic-level metal gaps and the effective number of molecules at hotspots. Firstly, immerse the gold substrate in the corresponding molecular solution, which can self-assemble molecular layers and achieve the deposition of a large number of molecules. Further use of the drop casting method to deposit gold nanoparticles, due to the extremely thin thickness of the molecular layer itself, makes it easy to achieve atomic-level metal gaps. The latest progress has enabled a large number of experiments using the NPoM structure, such as quantum heat engines~\cite{zhu2024autonomous}, low-frequency molecular vibration detection~\cite{boehmke2024uncovering}, and optoelectronic driven plasmonic nanocavity~\cite{doi:10.1126/sciadv.adt9808}. In experiments, the effective mode volume is typically $V_m<100$$\,$$\mathrm{nm}^{-3}$, which can be adjusted by employing metal nanoparticles of different sizes or shapes and altering the gap between nanoparticles and the gold mirror. Mechanical mode frequency is typically between $\omega_{b}/2\pi=0.749\sim48.3$$\,$$\mathrm{THz}$ and can also be adjusted by using different types of molecules. The Raman tensor typically ranges from $R_{b}=1.7\sim5840$$\,$${\AA}^{4}$amu$^{-1}$, and the optomechanical coupling coefficient in the range of $g=0.0295\sim9.7$$\,$$\mathrm{THz}$ can be achieved (more details see the Appendix.~\ref{A1}).
	
	In this article, a hybrid molecular COM system was constructed by combining the NPoM structure with a high-Q cavity. Under the current experimental conditions, it is completely feasible to integrate the two interactions into a unified system. At present, high-Q microcavity technology is very mature and can achieve $10^{10}$ or even higher. The latest developments have effectively combined traditional cavity COM with molecular COM. Various nanoparticle-on-cavity structures have emerged, such as WGM cavity, FP cavity, and photonic crystal cavity. In the case of the WGM cavity, the deposition of metal nanoparticles can induce back-scattering and give rise to multiple optical modes within the resonator. Photonic crystal cavities, meanwhile, often suffer from relatively lower quality factors. In contrast, the FP cavity can sustain a high quality factor while effectively mitigating back-scattering effects. In these experiments, the optical quality factor range of the microdisk cavity or FP cavity is $Q$$\,=\,$$10^2$$\,\sim\,$$10^6$, and optical coupling strength between the plasmonic cavity and microdisk cavity is within the range of $J/2\pi$$\,=\,$$0.1\sim10\,\mathrm{THz}$.

	\section{Conclusions}\label{sec4}
	In conclusion, we theoretically investigated the characteristics of the phonon laser in the proposed hybrid molecular COM system. Our findings demonstrate that the synergistic combination of exceptionally high single-photon optomechanical coupling strength and molecular collective effect enables the realization of ultra-low threshold phonon laser with $\mathrm{P}_{\mathrm{th}}=17.5\,\mathrm{nW}$, even under conditions of extremely low optical quality factor $Q_a=100$ and high mechanical damping rate $\gamma/2\pi=0.01\,\mathrm{THz}$. Moreover, the mechanical gain can be selectively enhanced or attenuated at different optical detuning, and the threshold power can be further reduced by optimizing the quality factor and adjusting the distance between mirrors in the FP cavity. More importantly, owing to the molecular characteristic $\omega_b/2\pi=20\,\mathrm{THz}$~\cite{benz2016single,esteban2022molecular,roelli2016molecular}, this device is capable of generating the MIR-frequency phonon laser. High-frequency phonon lasers are characterized by their superior resolution~\cite{RevModPhys.86.1391,10.1063/1.1499745}, strong penetration abilities~\cite{https://doi.org/10.1002/lpor.201000011}, and minimal noise levels~\cite{teufel2011sideband,chan2011laser}. \\ Beyond phonon lasing, molecular COM systems provide a versatile platform for exploring rich nonlinear and quantum phenomena. Future investigations may include quantum entanglement~\cite{PhysRevLett.98.030405,PhysRevLett.125.143605,PhysRevLett.112.080503,doi:10.1126/science.abb0328,PhysRevApplied.22.064001,PhysRevApplied.18.064008,liu2023phase}, photon blockade~\cite{faraon2008coherent,PhysRevLett.121.153601,rabl2011photon,nunnenkamp2011single}, and ultra-sensitive quantum sensing~\cite{gavartin2012hybrid,li2021cavity,Jing:18,wang2024quantum,zhang2024quantum,zhao2020weak,doi:10.1021/acs.nanolett.0c03119}. We hope and believe that THz phonon lasers can be applied in many scenarios, such as laser sensing~\cite{PhysRevApplied.16.044007,Pan.202400593,He:23}, 6G technology~\cite{Chafatinos2020,Li202100978}, biomedical imaging~\cite{doi:10.1073/pnas.2413938121,10.1063/1.3459142} and MIR acoustics.
	
	\begin{appendix}
		\renewcommand{\thesection}{Appendix}
		\appendix
		\section{MOLECULAR COM-RELATED PARAMETERS}\label{A1}
			Here, we summarize the structure and types of NPoM molecules in molecular COM [see Table~\ref{Table-2}]. We found that the effective mode volumes of these cavity fields are all below $V_m<100$$\,$$\mathrm{nm}^{-3}$. The resonance frequency of the molecule is between $\omega_{b}/2\pi=0.749\sim48.3$$\,$$\mathrm{THz}$. In addition, according to the calculations in the reference~\cite{roelli2016molecular}, the single-photon optomechanical coupling strength in the last four rows of the table is $g=0.0295\sim0.628$$\,$$\mathrm{THz}$ when the effective mode volume is $V_m=1500$$\,$$\mathrm{nm}^{-3}$ (Obviously, compared to the experiments in the table, this effective mode volume is larger) and the cavity field resonance frequency is $\lambda_a=900$$\,$$\mathrm{nm}$. Thus, from the table, we can find molecules similar to our work, such as R6G, GBT, graphene, or the G band of CNT.
		\begin{table}[htbp]
			\caption{Summary of key parameters for molecular COM}
			\label{Table-2}
			\centering
			\setlength{\tabcolsep}{2pt}  
			\renewcommand{\arraystretch}{1.6} 
			\setlength{\extrarowheight}{4pt} 
			\arrayrulecolor{black} 
			\begin{tabular}{ccccccc}
				\hline
				\makecell[c]{Types of \\ nanoparticles \\ (size/nm)} & \makecell[c]{Cavity resonance \\ wavelength/nm} & \makecell[c]{Effective mode \\ volume/$\mathrm{nm^3}$ \\ (Gap/nm)}  & \makecell[c]{Types of \\ molecules} & \makecell[c]{Molecular \\ Raman peak/$\mathrm{cm^{-1}}$} & \makecell[c]{Molecular \\ Raman intensity \\ /$\mathrm{\AA^4amu^{-1}}$} & \makecell[c]{Single-photon \\ optomechanical \\ coupling/$\mathrm{THz}$} \\
				\hline
				\makecell[c]{Au sphere  \\ (80)~\cite{J.J.BaumbergUncovering}} &  753-856 & (0.8-1.9) & \makecell[c]{Thiol \\ molecules} & 25-200,1080 &  & $>0.725$  \\ 
				\hline
				\makecell[c]{Au sphere  \\ (80)~\cite{J.J.BaumbergGiant}} & 520,670,830 & $<1$ (1.3) & BPT  & 1586 & 5840 & 1.2-9.7  \\ 
				\hline
				\makecell[c]{Au/Ag sphere  \\ (90)~\cite{J.J.BaumbergSingle-molecule}} & 683,809 & 22 (1.3) & BPT & 625-1511 & &  \\
				\hline
				\makecell[c]{Au/Ag sphere (55) \\ +$SiO_2$ shell (2)~\cite{J.J.BaumbergResolving}} & 560,710 &  & \makecell[c]{MG, \\ pNTP, \\ PIC}  & 1211-1610 & &  \\
				\hline
				\makecell[c]{Au sphere \\ (40)~\cite{Shell-isolated}} & 632.8 & 100 ($<2$) & \makecell[c]{NC-BPT, \\ BPT} & 1320,1500 & &  \\
				\hline
				\makecell[c]{Au sphere (60-100) \\ /Ag cube (65)~\cite{J.J.BaumbergStrong-coupling}} & 664,761,785,800 & $<1$ (0.7-10)  & \makecell[c]{$WSe_2,$ \\ $MoS_2$} & 250,257 & &  \\
				\hline
				\makecell[c]{Ag cube \\ (55.8)~\cite{Quantifying}} & 760,798 & (1.11)  & $MoS_2$ & 385,404 & &  \\
				\hline
				\makecell[c]{Au sphere \\ (80)~\cite{In-operando}} & 785 & (1.15)  & BPT & 1080,1585 & &  \\
				\hline
				\makecell[c]{Au sphere \\ (80)~\cite{J.J.BaumbergOptical}} & 633,785 & $<1$ (1.3)  & BPT & 290,1050 & & \\
				\hline
				\makecell[c]{Au cube \\ (85)~\cite{Plasmonic}} & 780 & (1.8) & $WSe_2$ & & &  \\
				\hline
				\makecell[c]{Au cube \\ (80)~\cite{Active}} & 615,751 & (2.5)  & $WSe_2$ &  &  & 1.3  \\
				\hline
				\makecell[c]{Au sphere \\ (100)~\cite{Acoustic}} & 550,670 & (4)  & \makecell[c]{PDMS, \\ MPTMS, \\ APTES} & 150,236,320 & &  \\
				\hline
				~\cite{DFTVibrational} &  &  & R6G & 1301,1351 & 5.9,351.7 &  \\
				\hline
				~\cite{Amultiscale} &  &  & Thiophenol & 9998,1072 & 31.6,1.7 &  \\
				\hline
				~\cite{Amultiscale} &  &  & GNT & 1000,1072 & 29.5,388.5 &  \\
				\hline
				~\cite{Absolute} &  &  & \makecell[c]{Graphene, \\ CNT} & 1600 & $10^3-10^4$ & \\
				\hline
			\end{tabular}
		\end{table}
		\section{DERIVATION OF THE HAMILTONIAN}\label{A2}
		Due to the collective effect of molecules $B=\sum_{i=1}^{N}b_{i}/\sqrt{N}$ ~\cite{PhysRevB.110.184306,zou2024amplifying}, the Hamiltonian of the system can be simplified as:
		\begin{eqnarray}
			\begin{aligned}
				H =&\hbar\omega_a a^{\dagger}a+\hbar\omega_c c^{\dagger}c+\hbar\omega_{b}B^{\dagger}B+\hbar J(a^{\dagger}c+ac^{\dagger}) \\ &-\hbar g_{N}a^{\dagger}a(B^{\dagger}+B)+i\hbar\epsilon_{l}(a^{\dagger}e^{-i\omega_l t}-ae^{i\omega_l t}). \\
			\end{aligned}
			\label{a1}
		\end{eqnarray}
		By using the unitary transformation $U=e^{-i\omega_{l}t(a^{\dagger}a+c^{\dagger}c)}$, the Hamiltonian of system can be simplified as:
		\begin{eqnarray}
			\begin{aligned}
				H =&-\hbar\Delta_a a^{\dagger}a-\hbar\Delta_c c^{\dagger}c+\hbar\omega_{b}B^{\dagger}B+\hbar J(a^{\dagger}c+ac^{\dagger}) \\ &-\hbar g_{N}a^{\dagger}a(B^{\dagger}+B)+i\hbar\epsilon_{l}(a^{\dagger}-a).
			\end{aligned}
			\label{a2}
		\end{eqnarray}
		Then we introduce the supermode operators $a_{\pm}=(a\pm c)/\sqrt{2}$ satisfying the relations,
		\begin{eqnarray}
			\begin{aligned}
				[a_{+},a_{+}^{\dagger}]=[a_{-},a_{-}^{\dagger}]=1,[a_{+},a_{-}^{\dagger}]=0.
			\end{aligned}
			\label{a3}
		\end{eqnarray}
		$H_{0}$ and $H_{\mathrm{dr}}$ can be written as 
		\begin{eqnarray}
			\begin{aligned}
				H_{0}=&\omega_{+}a_{+}^{\dagger}a_{+}+\omega_{-}a_{-}^{\dagger}a_{-}+\omega_{b}B^{\dagger}B,\\ H_{\mathrm{dr}}=&\frac{i\hbar}{\sqrt{2}}[\epsilon_{l}(a_{+}^{\dagger}+a_{-}^{\dagger})-\mathrm{H.c.}],
			\end{aligned}
			\label{a4}
		\end{eqnarray}
		with the supermode frequencies $\omega_{\pm}=-(\Delta_{a}+\Delta_{c})/2\pm J$.
		Under the rotating-wave-approximation condition $2J+\omega_{b},\omega_{b}\gg|2J-\omega_{b}|$~\cite{PhysRevApplied.10.064037,PhysRevA.103.053501}, we can obtain the interaction Hamiltonian:
		\begin{eqnarray}
			\begin{aligned}
				H_{\mathrm{I}}=&-\frac{\hbar g_{N}}{2}(a_{+}^{\dagger}a_{-}B+B^{\dagger}a_{-}^{\dagger}a_{+})+\frac{\hbar\Delta_{s}}{2}(a_{+}^{\dagger}a_{-}+a_{-}^{\dagger}a_{+}).
			\end{aligned}
			\label{a5}
		\end{eqnarray}

		\section{DERIVATION OF THE MECHANICAL GAIN}\label{A3}		
		In this supermode picture, the equations of motion can be written as
		\begin{eqnarray}
			\begin{aligned}
				\dot{a_{+}}=&-(i\omega_{+}+\kappa)a_{+}+\frac{i}{2}(g_{N}B-\Delta_{s})a_{-}+\frac{\epsilon_{l}}{\sqrt{2}}, \\ \dot{a_{-}}=&-(i\omega_{-}+\kappa)a_{-}+\frac{i}{2}(g_{N}B^{\dagger}-\Delta_{s})a_{+}+\frac{\epsilon_{l}}{\sqrt{2}}, \\ \dot{B}=&-(i\omega_{b}+\gamma)B+\frac{ig_{N}}{2}a_{-}^{\dagger}a_{+}.
			\end{aligned}
			\label{a7}
		\end{eqnarray} 
		Then we can define the ladder operator $p=a_{-}^{\dagger}a_{+}$ and the population inversion operator as $\delta n=a_{+}^{\dagger}a_{+}-a_{-}^{\dagger}a_{-}$. The equations of the system then read
		\begin{eqnarray}
			\begin{aligned}
				\dot{B}=&-(i\omega_{b}+\gamma)B+\frac{ig_{N}}{2}p, \\ \dot{p}=&-2(iJ+\kappa)p+\frac{i}{2}(\Delta_{s}-g_{N}B)\delta n+\frac{\epsilon_{l}}{\sqrt{2}}(a_{-}^{\dagger}+a_{+}).
			\end{aligned}
			\label{a8}
		\end{eqnarray} 
		Thus, we obtain the steady-state values, that is,
		\begin{eqnarray}
			\begin{aligned}
				a_{+}=&\frac{\epsilon_{l}(ig_{N}B-i\Delta_{s}+2i\omega_{-}+2\kappa)}{2\sqrt{2}[D-i\kappa(\Delta_{a}+\Delta_{c})]}, \\ a_{-}=&\frac{\epsilon_{l}(ig_{N}B^{\dagger}-i\Delta_{s}+2i\omega_{+}+2\kappa)}{2\sqrt{2}[D-i\kappa(\Delta_{a}+\Delta_{c})]}, \\ p=&\frac{\sqrt{2}\epsilon_{l}(a_{-}^{\dagger}+a_{+})-i(g_{N}B-\Delta_{s})\delta n}{2i(2J-\omega_{m})+4\kappa},
			\end{aligned}
			\label{a9}
		\end{eqnarray} 
		with
		\begin{eqnarray}
			\begin{aligned}
				D=&J^{2}+\kappa^{2}-\Delta_{a}\Delta_{c}+\frac{g_{N}^{2}B^{\dagger}B-2\Delta_{s}g_{N}Re(B)}{4}.
			\end{aligned}
			\label{a10}
		\end{eqnarray} 
		Substitution of Eq.~\ref{a9} into the dynamical equation of $B$ in Eq.~\ref{a8} results in
		\begin{eqnarray}
			\begin{aligned}
				\dot{B}=(-i\omega_{b}-i\omega^{\prime}+\mathrm{G}-\gamma)B+F,
			\end{aligned}
			\label{a11}
		\end{eqnarray} 
		where
		\begin{eqnarray}
			\begin{aligned}
				\omega^{\prime}=&\frac{(2J-\omega_{m})g_{N}^{2}\delta n}{4(2J-\omega_{m})^{2}+16\kappa^{2}}+\frac{g_{N}^{2}\epsilon_{l}^{2}\kappa^{2}(\Delta_{a}+\Delta_{c})}{[2(2J-\omega_{m})^{2}+8\kappa^{2}][D^{2}+\kappa^{2}(\Delta_{a}+\Delta_{c})^{2}]}, \\ F=&\frac{-g_{N}\Delta_{s}\delta n}{4i(2J-\omega_{m})+8\kappa}+\frac{i\epsilon_{l}^{2}g_{N}[D(\kappa-iJ)+\kappa(\Delta_{a}+\Delta_{c})\Delta_{c}]}{[2i(2J-\omega_{m})+4\kappa][D^{2}+\kappa^{2}(\Delta_{a}+\Delta_{c})^{2}]},
			\end{aligned}
			\label{a12}
		\end{eqnarray} 
		and the mechanical gain is
		\begin{eqnarray}
			\begin{aligned}
				\mathrm{G}=&\frac{g_{N}^{2}\kappa\delta n}{2(2J-\omega_{m})^{2}+8\kappa^{2}}+\frac{\epsilon_{l}^{2}g_{N}^{2}(\omega_{m}-2J)(\Delta_{a}+\Delta_{c})\kappa}{[4(2J-\omega_{m})^{2}+16\kappa^{2}][D^{2}+(\Delta_{a}+\Delta_{c})^{2}\kappa^{2}]},
			\end{aligned}
			\label{a13}
		\end{eqnarray} 
		where $\delta n$ can be expressed as
		\begin{eqnarray}
			\begin{aligned}
				\delta n=\frac{\epsilon_{l}^{2}[2J\Delta_{c}-Jg_{N}Re(B)-\kappa g_{N}Im(B)]}{D^{2}+\kappa^{2}(\Delta_{a}+\Delta_{c})^{2}}.
			\end{aligned}
			\label{a14}
		\end{eqnarray}

		\section{STATISTICAL PROPERTIES OF THE PHONON}\label{A4}
		Under the small fluctuation approximation as shown in
		\begin{eqnarray} 
			\begin{aligned}
				a_{+}(t)&=a_{1,s}+\Lambda_{1}(t),a_{-}(t)=a_{2,s}+\Lambda_{2}(t),B(t)=B_{s}+\beta(t),
			\end{aligned}
			\label{a15}
		\end{eqnarray}
		by substituting Eq.~\ref{a15} into Eq.~\ref{a7}, we can obtain the dynamic equations of fluctuation operators,
		\begin{eqnarray} 
			\begin{aligned}
				\dot{\Lambda_{1}(t)}=&-(i\omega_{+}+\kappa)\Lambda_{1}(t)-\frac{i}{2}\Delta_{s}\Lambda_{2}(t)+\sqrt{2\kappa}\Gamma_{1}(t)+\frac{i}{2}g_{N}[B_{s}\Lambda_{2}(t)+a_{2,s}\beta(t)], \\ \dot{\Lambda_{2}(t)}=&-(i\omega_{-}+\kappa)\Lambda_{2}(t)-\frac{i}{2}\Delta_{s}\Lambda_{1}(t)+\sqrt{2\kappa}\Gamma_{2}(t)+\frac{i}{2}g_{N}[B_{s}^{*}\Lambda_{1}(t)+a_{1,s}\beta^{*}(t)], \\\dot{\beta(t)}=&-(i\omega_{b}+\gamma)\beta(t)+\sqrt{2\gamma}B_{in}(t)+\frac{ig_{N}}{2}[a_{2,s}^{*}\Lambda_{1}(t)+a_{1,s}\Lambda_{2}^{*}(t)],
			\end{aligned}
			\label{a16}
		\end{eqnarray}
		where $\Gamma_{1}(t)$, $\Gamma_{2}(t)$ and $B_{in}(t)$ represent fluctuation operators corresponding to
		the optical and mechanical resonators. If we introduce the Fourier transform $f(t)=\int_{-\infty}^{+\infty}f(\omega)exp(-i\omega t)(d\omega/2\pi)$ for the arbitrary smooth function $f(t)$, the motion equations for the fluctuation operators in the
		frequency domain can be written as
		\begin{eqnarray} 
			\begin{aligned}
				-i\omega\Lambda_{1}(\omega)=&-(i\omega_{+}+\kappa)\Lambda_{1}(\omega)-\frac{i}{2}\Delta_{s}\Lambda_{2}(\omega) \\&+\frac{ig_{N}}{2}[B_{s}\Lambda_{2}(\omega)+a_{2,s}\beta(\omega)]+\sqrt{2\kappa}\Gamma_{1}(\omega),
				\\-i\omega\Lambda_{2}(\omega)=&-(i\omega_{-}+\kappa)\Lambda_{2}(\omega)-\frac{i}{2}\Delta_{s}\Lambda_{1}(\omega) \\&+\frac{ig_{N}}{2}[B_{s}^{*}\Lambda_{1}(\omega)+a_{1,s}\beta^{\dagger}(\omega)]+\sqrt{2\kappa}\Gamma_{2}(\omega), \\-i\omega\beta(\omega)=&-(i\omega_{b}+\gamma)\beta(\omega)+\sqrt{2\gamma}B_{in}(\omega)+\frac{ig_{N}}{2}[a_{2,s}^{*}\Lambda_{1}(\omega)+a_{1,s}\Lambda_{2}^{\dagger}(\omega)].
			\end{aligned}
			\label{a17}
		\end{eqnarray}
		By eliminating the fluctuation operators $\Lambda_{1}(\omega)$ and $\Lambda_{2}(\omega)$, we can obtain the expression of $\beta(\omega)$,
		\begin{eqnarray} 
			\begin{aligned}
				\beta(\omega)=&p_{1}b_{in}(\omega)+p_{2}b_{in}^{\dagger}(\omega)+p_{3}\Gamma_{1}(\omega)+p_{4}\Gamma_{1}^{\dagger}(\omega)+p_{5}\Gamma_{2}(\omega)+p_{6}\Gamma_{2}^{\dagger}(\omega),
			\end{aligned}
			\label{a18}
		\end{eqnarray}
		where 
		\begin{eqnarray} 
			\begin{aligned}
				\lambda_{1}^{\pm}&=\pm i\omega_{+}+\kappa-i\omega,\lambda_{2}^{\pm}=\pm i\omega_{-}+\kappa-i\omega, \\\lambda_{3}^{\pm}&=\pm i\omega_{b}+\gamma-i\omega,U^{\pm}=\lambda_{1}^{\pm}\lambda_{2}^{\pm}+\frac{1}{4}(g_{N}B_{s}-\Delta_{s})(g_{N}B_{s}^{*}-\Delta_{s}), \\ \mu_{1}&=-\frac{i}{8}g_{N}^{2}a_{1,s}a_{2,s}^{*}(g_{N}B_{s}-\Delta_{s})(U^{+}+U^{-}),\\ \mu_{2}&=\frac{i}{8}g_{N}^{2}a_{1,s}^{*}a_{2,s}(g_{N}B_{s}^{*}-\Delta_{s})(U^{+}+U^{-}), \\\Pi_{1}&=\lambda_{3}^{+}U^{+}U^{-}+\frac{1}{4}g_{N}^{2}|a_{2,s}|^{2}\lambda_{2}^{+}U^{-}-\frac{1}{4}g_{N}^{2}|a_{1,s}|^{2}\lambda_{1}^{-}U^{+}, \\\Pi_{2}&=\lambda_{3}^{-}U^{+}U^{-}+\frac{1}{4}g_{N}^{2}|a_{2,s}|^{2}\lambda_{2}^{-}U^{+}-\frac{1}{4}g_{N}^{2}|a_{1,s}|^{2}\lambda_{1}^{+}U^{-}, \\D&=\Pi_{1}\Pi_{2}-\mu_{1}\mu_{2},
			\end{aligned}
			\label{a19}
		\end{eqnarray}
		and
		\begin{eqnarray} 
			\begin{aligned}
				p_{1}&=U^{+}U^{-}\Pi_{2}\sqrt{2\gamma}/D,p_{2}=U^{+}U^{-}\mu_{1}\sqrt{2\gamma}/D, \\ p_{3}&=[\frac{1}{4}g_{N}a_{1,s}^{*}(g_{N}B_{s}^{*}-\Delta_{s})U^{-}\mu_{1}+\frac{i}{2}g_{N}a_{2,s}^{*}\lambda_{2}^{+}U^{-}\Pi_{2}]\sqrt{2\kappa}/D, \\p_{4}&=[\frac{1}{4}g_{N}a_{1,s}(g_{N}B_{s}-\Delta_{s})U^{+}\Pi_{2}-\frac{i}{2}g_{N}a_{2,s}\lambda_{2}^{-}U^{+}\mu_{1}]\sqrt{2\kappa}/D, \\p_{5}&=[-\frac{1}{4}g_{N}a_{2,s}^{*}(g_{N}B_{s}-\Delta_{s})U^{-}\Pi_{2}-\frac{i}{2}g_{N}a_{1,s}^{*}\lambda_{1}^{+}U^{-}\mu_{1}]\sqrt{2\kappa}/D, \\p_{6}&=[-\frac{1}{4}g_{N}a_{2,s}(g_{N}B_{s}^{*}-\Delta_{s})U^{+}\mu_{1}+\frac{i}{2}g_{N}a_{1,s}\lambda_{1}^{-}U^{+}\Pi_{2}]\sqrt{2\kappa}/D.
			\end{aligned}
			\label{a19}
		\end{eqnarray}
		Then, we obtain the correlation functions of the phonon fluctuation operators in the frequency domain as follows:
		\begin{eqnarray} 
			\begin{aligned}
				\langle\beta^{\dagger}(\omega)\beta(\omega^{\prime})\rangle&=2\pi X_{\beta^{\dagger}\beta}\delta(\omega+\omega^{\prime}), \\ \langle\beta(\omega)\beta(\omega^{\prime})\rangle&=2\pi X_{\beta\beta}\delta(\omega+\omega^{\prime}),
			\end{aligned}
			\label{a20}
		\end{eqnarray}
		with 
		\begin{eqnarray} 
			\begin{aligned}
				X_{\beta^{\dagger}\beta}(\omega)=&|p_{1}(-\omega)|^{2}n_{b}+|p_{2}(-\omega)|^{2}(n_{b}+1)+|p_{4}(-\omega)|^{2}+|p_{6}(-\omega)|^{2}, \\X_{\beta\beta}(\omega)=&p_{1}(\omega)p_{2}(-\omega)(n_{b}+1)+p_{2}(\omega)p_{1}(-\omega)n_{b}+p_{3}(\omega)p_{4}(-\omega)+p_{5}(\omega)p_{6}(-\omega).
			\end{aligned}
			\label{a21}
		\end{eqnarray}
		Here, $X_{\beta^{\dagger}\beta}(\omega)$ and $X_{\beta\beta}(\omega)$ represent the correlation spectra of the phonon fluctuation operators, and $n_{b}=1\big/\left(e^{\hbar\omega_{b}/(k_{B}T)}-1\right)$ is the average thermal phonon occupation number. Therefore, different temperatures $T$ result in different thermal phonon occupation numbers and correlation spectra. If we assume that the environmental noises have Gaussian distributions, then, from Wick’s theorem, with straightforward but tedious calculations, we have $\langle\beta^{\dagger}(t)\beta^{\dagger}(t)\beta(t)\beta(t)\rangle=2Y_{\beta^{\dagger}\beta}^{2}+|Y_{\beta\beta}|^{2}$, where $Y_{\beta^{\dagger}\beta}$ and $Y_{\beta\beta}$ can be calculated as
		\begin{eqnarray} 
			\begin{aligned}
				Y_{\beta^{\dagger}\beta}&=\langle\beta^{\dagger}(t)\beta(t)\rangle=\frac{1}{2\pi}\int_{-\infty}^{+\infty}X_{\beta^{\dagger}\beta}(\omega)d\omega, \\Y_{\beta\beta}&=\langle\beta(t) \beta(t)\rangle=\frac{1}{2\pi}\int_{-\infty}^{+\infty}X_{\beta\beta}(\omega)d\omega.
			\end{aligned}
			\label{a22}
		\end{eqnarray}
		Therefore, we can calculate the normalized equal-time second-order correlation function of phonons 
		\begin{eqnarray} 
			\begin{aligned}
				g^{(2)}(0)=&\frac{|B_{s}|^{4}+2Re[B_{s}^{*2}\langle\beta(t) \beta(t)\rangle]}{[|B_{s}|^{2}+\langle\beta^{\dagger}(t)\beta(t)\rangle]^{2}}+\frac{4|B_{s}|^{2}\langle\beta^{\dagger}(t)\beta(t)\rangle+\langle\beta^{\dagger}(t)\beta^{\dagger}(t)\beta(t)\beta(t)\rangle}{[|B_{s}|^{2}+\langle\beta^{\dagger}(t)\beta(t)\rangle]^{2}}.
			\end{aligned}
			\label{a23}
		\end{eqnarray}
	\end{appendix}
	
	\medskip
	\textbf{Acknowledgements} \par 
	This work is supported by the National Key R$\&$D Program of China (No. 2024YFE0102400), the National Natural Science Foundation of China (Grant No. 12421005, 12565001, 12205054), the Hunan Major Sci-Tech Program (2023ZJ1010), the Innovation Program for Quantum Science and Technology \\(2024ZD0301000), and the Natural Science Foundation of Jiangxi Province (20252BAC200163).
	
	\textbf{Conflicts of Interest} \par
	The authors declare no conflicts of interest.
	
	\textbf{Data Availability Statement} \par 
	Data sharing not applicable to this article as no datasets were generated or analysed during the current study.
	
	\medskip
	
	%
	


\end{document}